\documentclass{aa}
\usepackage{graphicx}
\begin{document}

   \title{Three component model of cosmic ray spectra from 10 GeV  to 100 PeV }

   \author{V.I. Zatsepin
          \and
          N.V. Sokolskaya
          }


   \institute{Skobeltsyn Institute of Nuclear Physics, Moscow State University,
              Russia\\
              \email{viza@dec1.sinp.msu.ru}
         }


  \abstract
   {}
   {
    A model to describe cosmic ray spectra in the energy region from $10^{10}$
     to $10^{17}$ eV is suggested based on the assumption that Galactic cosmic ray flux
    is a mixture of  fluxes accelerated by shocks from nova and supernova  of
    different types.
   }
   {
     We analyze  recent experimental data on cosmic ray spectra obtained in direct measurements
     above the atmosphere and data obtained with ground Extensive Air Shower arrays.
   }
   { The model of the three classes of cosmic ray sources  is consistent with
   direct experimental data on cosmic ray elemental spectra and gives a smooth transition
   from  the all particle spectrum measured in the direct experiments to the all particle spectrum
   measured with EAS.}
   {}


\titlerunning{Three component model...}
    \maketitle
%

\section{Introduction}

   It is thought that cosmic rays below $10^{17}$ eV are accelerated by  shocks
   of supernova explosions, with all components having the same rigidity spectra.
   However,  recent experimental data show that spectral indices of elemental spectra
   are different.
    Therefore  we suggest  that the concepti of a single  population of  sources
   with the same index of rigidity spectra for all elements
    is inaccurate
   We  describe the experimental data
   using the idea that the measured cosmic ray flux is generated  by three different
   classes of  sources. It is suggested that each class of sources generates the spectrum
   that is  power-law by rigidity  with its specific spectral index and
   maximal rigidity.
    We  analyze  cosmic ray spectra, measured by  direct methods, and  all-particle spectra,
   measured in the extensive air showers (EAS). The data used are shown
   in Table 1.

\begin{table*}
\caption{Experiments used in this paper}             
\centering                          
\begin{tabular}{l l l l l}        
\hline\hline                 
Experiment & Technique &Site & Reference \\    
\hline                        

AMS & Magnetic spectrometer & Spacecraft&  Alcaraz et al., \cite{ams} \\
CAPRICE & Magnetic spectrometer & Balloon&  Boezio et al., \cite{caprice} \\
BESS-TEV& Magnetic spectrometer & Balloon&  Haino et al., \cite{caprice} \\
ATIC-2 & Calorimeter & Balloon & Wefel et al., \cite{atic}; Panov et al., \cite{panov} \\
SOKOL&Calorimeter & Spacecraft&  Ivanenko et al., \cite{sokol} \\
JACEE& Emulsion chamber & Balloon&  Asakimori et al., \cite{jacee}; Takahashi et al., \cite{jacee_nucl} \\
MUBEE& Emulsion chamber& Balloon& Zatsepin et al.,\cite{zatsepin1}; \cite{mubee}\\
RUNJOB& Emulsion chamber& Balloon&  Derbina et al., \cite{runjob} \\
HEAO& \t Cerenkov counter& Spacecraft&  Engelmann et al., \cite{heao} \\
CRN& Transition radiation & Spacecraft&  M\"uller et al., \cite{crn} \\
TRACER& \t Transition radiation& Balloon&  M\"uller et al., \cite{tracer} \\
TIC& Calorimeter & Balloon&  Adams et al., \cite{tic} \\
KASCADE& EAS& Ground based&  Roth et al., \cite{kascade} \\
HEGRA-AIROBIC& EAS+\t Cerenkov light&Ground based& Arqueros et al.,\cite{hegra}\\
CASA-BLANCA& EAS+ \t Cerenkov light& Ground based&  Fowler al., \cite{casa-blanca} \\
DICE& \t Cerenkov light& Ground based&  Kieda et al., \cite{dice} \\
TUNKA& \t Cerenkov light& Ground based&  Budnev et al., \cite{tunka} \\
\hline                                   
\end{tabular}
\end{table*}

\section{Basis for the model}
   \begin{figure}
   \centering
   \includegraphics[width=8cm]{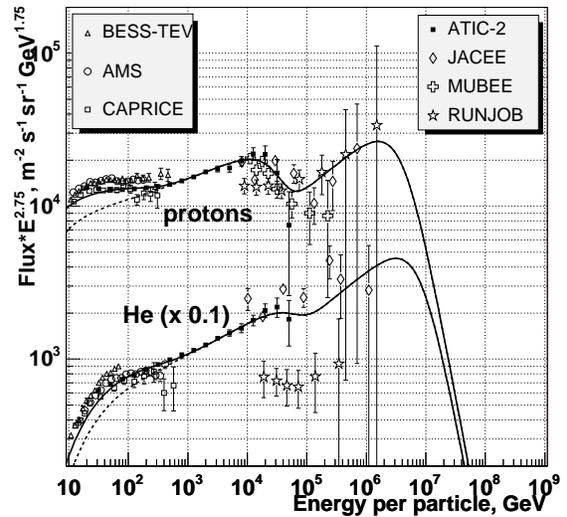}
      \caption{
                Proton  and He  spectra.
      Dashed lines are  described in Section 3, solid lines are described in Section 5.
              }
   \end{figure}
   The model is based on the preliminary experimental data obtained in the ATIC-2
   experiment
   (Wefel et al, \cite{atic}; Panov et al, \cite{panov}; Batkov et al, \cite{batkov})
   that  fill the gap between  measurements with magnetic spectrometers below 100 GeV
   (Haino et al, \cite {bess},
   Alcaraz et al, \cite{ams}, Boezio et al, \cite{caprice}) and  emulsion experiments above 10 TeV
   (Zatsepin et al, \cite{mubee}, Asakimori et al, \cite{jacee},
   Derbina et al, \cite{runjob}).
   The proton and helium spectra obtained in these direct measurements  are shown in Fig.1.
   The preliminary ATIC-2 data  show
   that the proton  spectrum obeys a power law with  a spectral index of $\gamma = 2.63\pm 0.01$
   in the energy region between $3\times 10^2$  and $10^4$ GeV .
   However, the extrapolation of this spectrum to  higher energy does not agree with the emulsion
   experiment data. Let us suppose that this difference is real, and that we see the end of acceleration
    for the first class of sources, while sources of the second class accelerate cosmic rays
    efficiently up to the knee region at about $3\times 10^{15}$ eV. Initially  we  suppose that
     the steeper spectrum below $3\times 10^2$ GeV is connected with weak reacceleration of cosmic rays
    during their propagation from the sources to the Earth.
    The helium spectrum  is flatter than the proton one. This difference may be explained by the
    assumption that the second type of sources has a different spectral index, and that the elemental
    composition for these two classes of sources are different.

\section {Description of the model}

The equilibrium spectra in the Galaxy will be described in the framework of the Leaky Box model:
$$ I(R) =\frac{ Q(R) \times \tau_{esc} (R) }{ 1+ \lambda_{esc}(R) / \lambda_p}, \eqno (1)$$
where R is the particle rigidity, $Q(R)$ is the source spectrum, $\tau_{esc}$ is the cosmic ray
life time
for the escape, $\lambda_p$ is the particle mean free path for interaction, and $\lambda_{esc}$ is
the escape length
($\lambda_{esc}(R) = \rho \times v \times\tau_{esc}(R),$ where $\rho$ is the density
of the interstellar medium and $v$ is the particle velocity).

The Leaky Box model without reacceleration adequately describes  the  spectra of nuclei and the
ratios of secondary nuclei to primary ones in the HEAO-3-C2 experiment below
35 GeV/nucleon (Engelmann et al,\cite{heao}),
if
 $$ \lambda_{esc}(R) = 34.1 \times R^{-0.6} \mbox{ g/cm$^2$}, \mbox{ for } R >= 4.4 \mbox{ GV}\eqno(2)$$
$$ \lambda_{esc}(R) =  10.8 \times \beta, \mbox{ g/cm$^2$ for } R < 4.4 \mbox{ GV}$$
$$\mbox{and }Q(R) \sim R^{-2.23};$$
but we  assume that at high energy $ \lambda_{esc}(R) \sim R^{-1/3} $, that  corresponds to the
Kolmogorov type of magnetic turbulence, and that the dependence  in eq.(2)  at $R < 100$ GV
results from reacceleration of cosmic rays during their propagation in the Galaxy.
According to Osborne and Ptuskin (\cite{osb_ptuskin}) reacceleration may be taken into account as
follows:
   $$ \lambda_{esc}(R) = 4.2 \times (R/R_0)^{-1/3} \times [1+(R/R_0)^{-2/3}] \mbox { g/cm$^2$}, \eqno (3)$$
   where $R_0 = 5.5$ GV.
   It is assumed that $$Q(R) \sim R^{-\alpha} \times \phi(R), \eqno (4)$$
   where $\alpha$ is the index of the source spectrum,
   and the function $\phi(R)$ describes smooth transition from the
   spectral index   in the region of effective acceleration for each type of source to the
   spectral    index after termination of  this process.
   $$ \phi(R) = [1+(R/R_{max})^2]^{(\gamma -\gamma_k)/2} \eqno (5)$$
   where $\gamma = \alpha+0.33$ is the spectral index in the region of effective acceleration (at high enough energy),
   and $\gamma_k$ is the spectral index after termination of  effective acceleration.
   It is assumed that spectra are simple power laws with  the index $\gamma_k$ after termination of
   effective acceleration.

  As later we will fit spectra of nuclear groups and the all particle spectrum,
  we convert  rigidity spectra to spectra by   energy per particle $E$.
  $$ I(E) = \frac{Q_p(R)\times \tau_{esc}(R)}{1+\lambda_{esc}(R)/\lambda_p}\times \frac{dR}{dE} \eqno (6)$$

   $$R =\frac{1}{Z} \times \sqrt{E^2 +2m_p\times A\times E } \mbox{ ;   }$$
    $$\frac{dR}{dE} = \frac{1}{Z} \times \frac{E + m_p A}{\sqrt{E^2 +2 m_p\times A\times E}} \eqno (7)$$
   where $Z$ and $A$ are particle charge and atomic weight, and $m_{p}$ is proton mass.

The spectrum of each cosmic ray nuclear group is the sum of spectra from the
different  classes of sources.

We described sources of  class I with values of $\alpha = 2.3$ and R$_{max}$ = 50 TV,
and sources of class II  with  $\alpha = 2.1$ and R$_{max}$ = 4 PV. The intensities for various cosmic
ray groups were chosen to fit both the data of direct measurements
and the data of EAS in the high energy region.
The predictions of the model along with experimental data for  cosmic ray groups and the
all particle spectrum are shown in Figures 1-3 with dashed lines (above $10^4$ GeV this line
follows the solid line, which is described in Section 5).
The parameters of the model are shown in Tables 1 and 2.

  \begin{figure}
   \centering
   \includegraphics[width=8 cm]{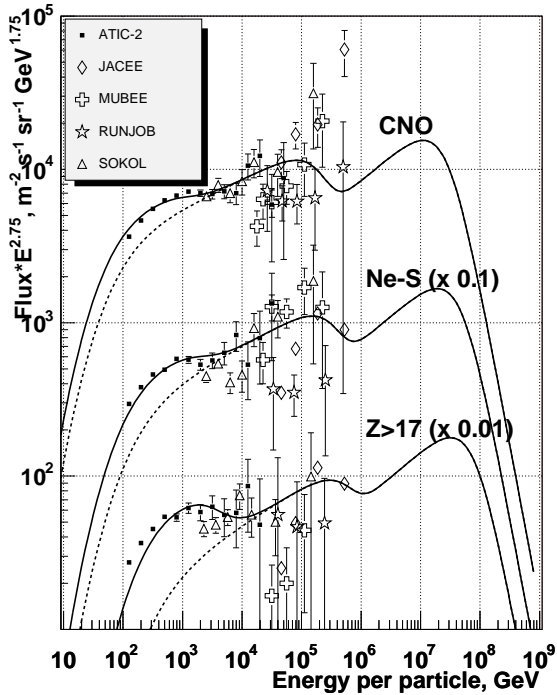}
   \caption{Spectra of nuclear groups.
   Dashed lines are  described in Section 3, solid lines are described in Section 5.}
   \end{figure}

  \begin{figure}
   \centering
   \includegraphics[width=8 cm]{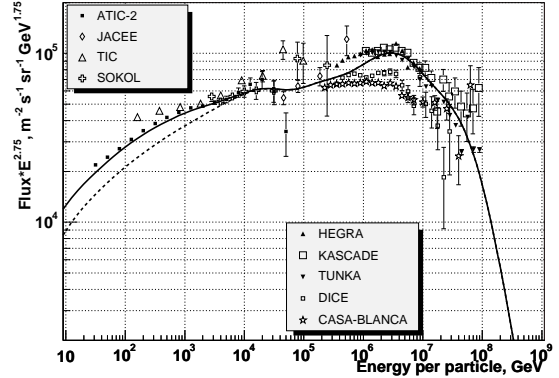}
   \caption{All particle spectrum.
   The dashed line is described in Section 3, the solid line is described in Section 5.}
   \end{figure}

 Above 300 GeV the proton  spectrum (Fig.1) is fitted closely by the model.
 The steepening of the
 helium spectrum above 10 TeV is not as clear as for proton one,  but the line does not
 contradict the experimental data.  The RUNJOB data for helium are much lower than the
 data of ATIC-2 and JACEE. This also holds true for other  groups
  of nuclei.
  In the region below $\sim 300$ GeV the model does not agree with the
 experimental data. An identical behavior is seen for the spectra of
 heavier nuclear groups shown  in Fig. 2 and for the all-particle spectrum shown in Fig.3.
  We will discuss this discrepancy in Section 5.
  The parameters for class II that determine
 the knee region  were chosen to fit the experimental data of HEGRA-AIROBICC, KASCADE and TUNKA,
 while the data of CASA-BLANKA and DICE appear to be below the model line.

  \section{Interpretation of the model}
 Our analysis is not the first attempt to link data of  direct measurements with data
 of EAS and to fit  the knee.
  Some previous models are  phenomenological, e.g. the poligonato model of H\"orandel
  (\cite {hoerandel})
   and some are based on physical grounds. In a number of papers, the origin of the knee is
   explained by the structure of
 magnetic fields in the Galaxy, or the change of inelastic
  interaction characteristics above several PeV.
   Erlykin and Wolfendale (\cite{erlykin})  assumed that the knee  arises from a single nearby
   cosmic ray  source. A detailed review of various models was made by H\"orandel (\cite{hoerandel_a}).

 As two classes of cosmic ray sources are needed for our model, we use the model of
 Bierman ( \cite{bierman}) where  two types of astrophysical objects
 are supposed as  sources of the Galactic cosmic rays. The first type is stars with
 masses from 8 to 15 of solar mass that exploded into the interstellar medium (ISM).
 The expected chemical composition generated by these objects is close to the composition of
 the ISM. The second type is massive stars exploding into their own stellar wind. The stars of 15 - 25 solar masses known as
 Red Supergiants (RSG) possess a moderate stellar wind and the stars of mass greater than 25 solar masses known as
 Wolf-Rayet (WR) stars have a strong stellar wind.
   The wind  stars are the late stage of evolution of massive O and B stars,
 which are clustered into
 OB associations. The combined effect of collective stellar winds and supernovae within
 OB associations  leads to the production of a superbubble
 (Heiles \cite{heiles}, Bruhweiler et al, \cite{bruhweiler}, Tomisaka \cite{tomisaka}), i.e.
 a huge (several hundred parsec) region of hot ($\sim 10^6$ K), low-density
 ($\sim 10^{-3}$ cm$^{-3}$) gas, surrounded by a cold  neutral hydrogen  supershell.
 It was supposed that the Sun is located within such
 a superbubble (Kafatos et al., \cite {kafatos}).
 It was shown (Kafatos et al.,\cite{kafatos}, Streitmatter et al., \cite{streitmatter},
 Lingenfelter et al.,\cite{lingenfelter}, Higdon and Lingenfelter, \cite{higdon},
 Higdon and Lingenfelter, \cite{higdon2}, Streitmatter and Jones \cite{streitmatter1})
 that in this case many observed features of
 cosmic rays  can be explained (age, anisotropy, some features of isotopic composition).
 We assume that our first class comes from with the explosions of isolated stars into the
 ISM, and that our second class comes from supernovae within the local superbubble.

 We assume that the second type of source generates cosmic rays with
  flatter spectra and with higher maximal energy.
 These are those cosmic rays that are studied
 by ground EAS arrays, and  that produce the knee in the PeV energy region.

 \section {Energy spectra below  $\sim $ 300 GeV per nucleon }
    \begin{figure}
   \centering
   \includegraphics[width=8 cm]{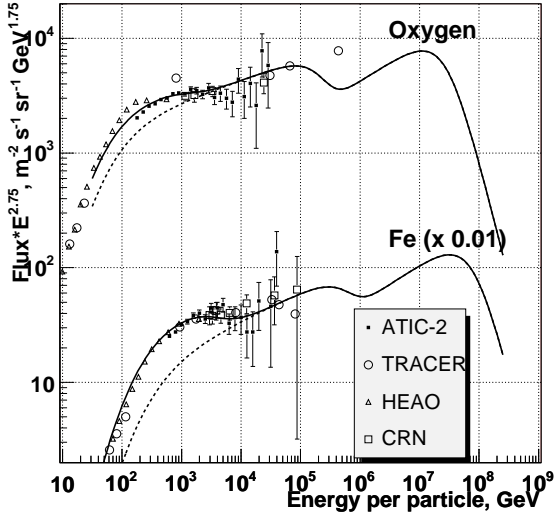}
 \caption{Spectra of oxygen and iron}
\end{figure}
\begin{figure}
   \centering
   \includegraphics[width=8 cm]{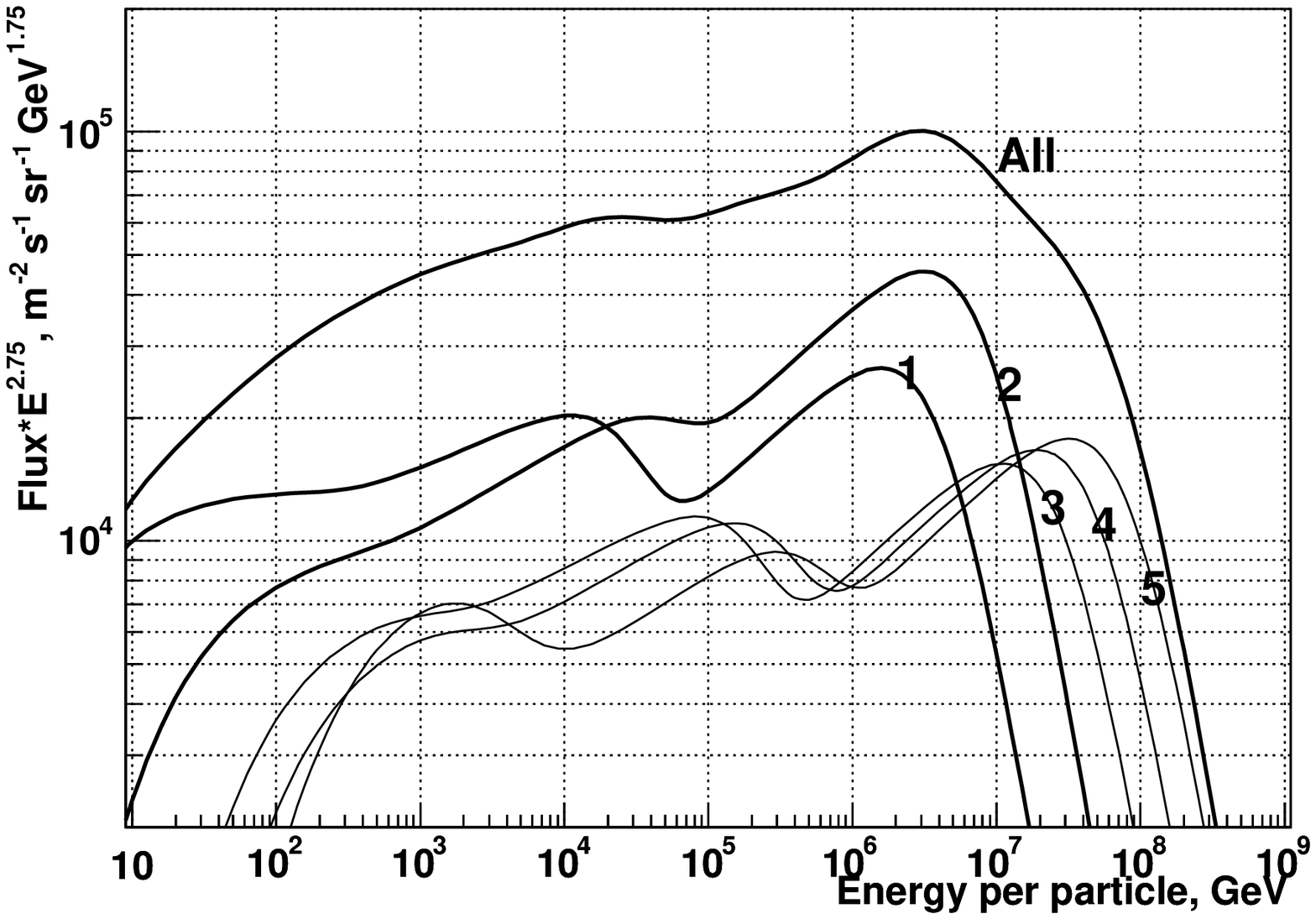}
   \includegraphics[width=8 cm]{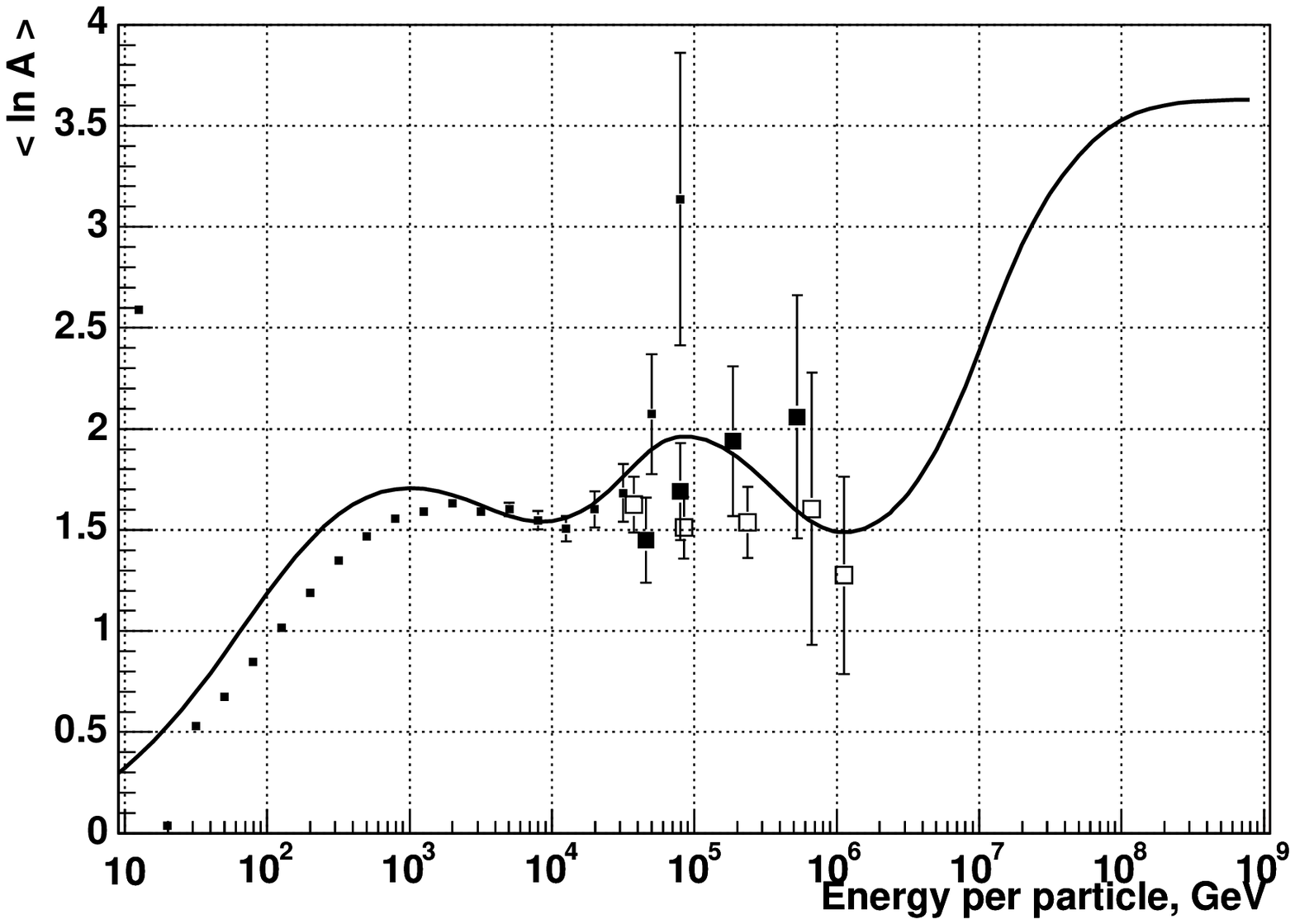}
    \caption{Top panel: Fluxes of cosmic ray groups in the model.  Line numbers 1: protons, 2: Helium,
   3: CNO group, 4:  Ne - S group, 5: Fe group (Z$>17$). Bottom panel: $<lnA>$ vs energy;
   small squares: preliminary ATIC-2, large solid squares: JACEE, open squares: RUNJOB}
\end{figure}
We now direct our attention to the energy region below $\sim 300$ GeV per nucleon. As one can see
 from Fig.1 - 3 (dashed lines),
 the model with two classes of sources does not fit  the preliminary ATIC-2 experimental data well.
 The weak reacceleration  is unable to fit the spectra in the low energy region and
 it may be suggested that most of  cosmic rays below 300 GeV per nucleon are
 accelerated in sources of a third class.
 The parameters for sources of the third class required to fit the experimental data are the following:
 $\alpha = 2.57;\mbox { } R_{max} = 200\mbox { } GV;\mbox{ } \gamma_k = 4.5.$
 In the region of such low rigidities, the solar modulation must be taken into account. We have used
 the eq. from Boezio( \cite{boezio}) for  this purpose with modulation parameter $\Phi = 600$ MV.
 $$ flux_{mod}(E) = flux(E+Ze\times\Phi)\times P,  \eqno (8)$$
 $$P =\frac{E^2+2m_pE}{(E+Ze\times \Phi)^2 +2m_p(E+Ze\times \Phi)},$$

 where  $e$ is the charge of the electron, $m_p$ is the proton mass and E is the kinetic energy of the nucleus.

 The intensities for various cosmic ray components
  were chosen  to fit the experimental data in Figures 1 - 3.
  The fit in the version of model with three classes of sources is shown in Fig. 1 - 3
  with solid lines.

  To check our model in this low energy region we plot  experimental data for separate heavy elements,
  oxygen and iron,  obtained from HEAO-3-C2 (Engelmann et al., \cite {heao}),
  CRN (M\"uller et al., \cite{crn}), and preliminary TRACER results (M\"uller et al., \cite{tracer})
  and the preliminary ATIC-2 data (Panov et al., \cite{panov}) along with two versions of the model
  in Fig.4.  The model of three classes of sources fits the experimental data well.
  The model also fits other individual spectra.

   Nova stars are candidates for the
  third class of sources. During nova explosion, an expanding shell  is produced, and the broad
  features of nova explosion are similar to these of the explosion of supernovae. Power requirements on the power are
   also satisfied: the mean energy of  explosion is $W \sim 10^{46} - 10^{47}$ erg.,  the
   frequency  of nova explosions is $\sim 100$ year$^{-1}$ ( Pskovsky, \cite{pskovsky}).
   So, the power available for cosmic
   ray generation is:
   $$\frac{(10^{46} - 10^{47})\times 100}{ 3\times 10^7} \approx 3\times 10^{40} -3\times 10^{41}\mbox{ erg/sec}$$
   It is not lower than the power supplied by supernova. Thus, this source class could generate
   cosmic rays with a   steeper spectrum ($\alpha = 2.57$) and of significantly lower maximal rigidity
   ( 200 GV).

\section{Parameters of the model, fluxes of different cosmic ray groups and $<$lnA$>$.}

 The  parameters  of the model used to fit spectra of cosmic ray groups are shown in Table 1
and Table 2.
Fig.5 shows fluxes of cosmic ray groups and the mean value of $<$lnA$>$ in the model.
The special feature of our model
 is that  charge composition is not a monotonic function of energy: it becomes 'lighter' before the
 knee, and only above the knee becomes heavier.

\begin{table}
\caption{
Parameters for three classes of sources. Class I: SN into ISM; class II:
SN in the Superbubble,
 class III: Novae.
 }             
\centering                          
\begin{tabular}{l l l l l}        
\hline                 
Class & $\alpha$ & $R_{max}[GV]$ & $\gamma$ & $\gamma_k$\\    
\hline                        
 I  & 2.3 & $5\times 10^4$ & 2.63 & 8 \\      
 II  & 2.1 & $4\times 10^6$ & 2.43 & 4.5 \\
 III & 2.57& $2\times 10^2$& 2.9& 4.5\\
\hline                                   
\end{tabular}
\end{table}
%
\begin{table}
\caption{ Flux $\times E^{2.75}$ [m$^{-2}$sec$^{-1}$ster$^{-1}$ GeV$^{1.75}$ for cosmic ray groups at E = 10 TeV }
\centering                          
\begin{tabular}{|l| c| c| c| c| c|}        
\hline

 & $Z_{eff}$&\multicolumn{3}{|c|}{Flux$\times E^{2.75}$}\\
\hline
&($\lambda_p$)&  class I &  class II&   class III\\    
\hline                        
 H & 1(74) &$1.4\times 10^4$ & $6.25 \times 10^3$ & 0.6 \\      
 He & 2(18) &$8.5\times 10^3$ & $8.5\times 10^3$ & 1.5 \\
 CNO & 7(5.8) &$6.75 \times 10^3$ & $1.8 \times 10^3$ &30 \\
 Ne-S & 12(3.5)& $5.5\times 10^3$ & $1.5\times 10^3$ & 110 \\
 $Z>17$ & 20(2.4)& $3.5\times 10^3$ & $1.2\times10^3$ & 750 \\
\hline                                   
\end{tabular}
\end{table}

\section{Conclusion}

 We proceed from the assumption that the difference between proton and helium spectral indices
 found in the JACEE experiment above 10 TeV (Asakimori et al.\cite{jacee}), and supported by the
 preliminary ATIC-2
 analysis (Wefel et al., \cite{atic})
 in the energy region above 100 GeV is real. We  suppose  that this difference results
 from the fact that  cosmic ray fluxes  are a mixture of fluxes from sources with different spectral
 indices and different maximal energy. We checked a simplest model assuming two classes of sources.
 We showed that elemental cosmic ray spectra pointed  to a 'kink' at the rigidity of about 50 TV,
 and connected this with the assumption that one class of sources terminates its effective
 acceleration  at this rigidity.
  This source class may be identified with  the supernova explosions into the
  ISM.  The second source class, presumably supernovae within the local
  superbubble, accelerates cosmic rays up to rigidity of 4 PV, producing the classic knee.
     However, the assumption of these two classes of sources is inadequate
 to fit the energy spectra below 300 GeV per nucleon along with the spectra in the high
 energy region, even taking into account minimal reacceleration of cosmic rays in the Galaxy.
 We assume that  the contribution  of nova stars is  essential in the energy region below
  $\sim$ 300 GeV per nucleon.
 \begin{acknowledgements}
 This work was supported by the Russian Foundation for Basic Research,
 grant number 05-02-16222. We are grateful to J.P. Wefel and T.G. Guzik for useful discussions.
\end{acknowledgements}

\end{document}